\documentclass[12pt,preprint,fleqn]{aastex}
\usepackage{graphicx}
\slugcomment{Accepted for publication in {\it The Astrophysical Journal}}

\begin{document}

\title{A Hadronic Synchrotron Mirror Model for the ``orphan'' 
TeV flare in 1ES 1959+650}

\author{Markus B\"ottcher\altaffilmark{1}}

\altaffiltext{1}{Astrophysical Institute, Department of Physics 
and Astronomy, Ohio University, Athens, OH 45701, USA}

\begin{abstract}
Very-high-energy $\gamma$-ray flares of TeV blazars are generally
accompanied by simultaneous flaring activity in X-rays. The
recent observations by the Whipple collaboration of an ``orphan''
TeV flare of 1ES~1959+650 (without simultaneous X-ray flare) is
very hard to reconcile with the standard leptonic SSC model which
is routinely
very successfully employed to explain the SED and
spectral variability of TeV blazars. In this paper, an alternative 
scenario is suggested in which the ``orphan'' TeV flare may
originate from relativistic protons, interacting with an external
photon field supplied by electron-synchrotron radiation reflected
off a dilute reflector. While the external photons will be 
virtually ``invisible'' to the co-moving ultrarelativistic
electrons in the jet due to Klein-Nishina effects, their Doppler
boosted energy is high enough to excite the $\Delta$ resonance 
from relativistic protons with Lorentz factors of $\gamma_p \sim
10^3$ -- $10^4$. This model is capable of explaining the ``orphan''
TeV flare of 1ES~1959+650 with plausible parameters, thus constraining
the number and characteristic energy of relativistic protons in the 
jet of this blazar.
\end{abstract}

\keywords{galaxies: active --- BL Lacertae objects: individual
(1ES~1959+650) --- gamma-rays: theory --- radiation mechanisms:
non-thermal}

\section{Introduction}

Blazars are a peculiar class of active galactic nuclei, consisting
of optically violently variable (OVV), gamma-ray loud quasars and
BL Lac objects. They have been observed at all wavelengths, from 
radio through very-high energy (VHE) $\gamma$-rays. Six 
blazars (Mrk~421: \cite{punch92}; Mrk~501: \cite{quinn96}; 
PKS 2155-314: \cite{chadwick99}; 1ES~2344+514: \cite{catanese98};
1H~1426+428: \cite{horan02}; 1ES~1959+650: \cite{kajino99,holder03})
have now been detected at VHE $\gamma$-rays ($> 350$~GeV) by 
ground-based air \v Cerenkov telescopes. Blazars exhibit variability 
at all wavelengths on various time scales. Radio interferometry 
often reveals one-sided kpc-scale jets with apparent superluminal 
motion.
The high inferred isotropic luminosities, short variability 
time scales, and superluminal motion provide conclusive evidence 
that
blazars are sources of relativistic jets pointing at a small 
angle with respect to our line of sight. 

One of the key unresolved questions in the field of blazar research 
to date is the nature of relativistic particles in blazar jets.
In the framework of relativistic jet models, the low-frequency 
(radio
-- optical/UV) emission from blazars is interpreted as 
synchrotron
emission from nonthermal relativistic electrons in the 
jet. The
high-frequency (X-ray -- $\gamma$-ray) emission could either 
be produced via Compton upscattering of low frequency radiation by the
same electrons responsible for the synchrotron emission \citep[for 
a recent review see, e.g.,][]{boettcher02}, or due to hadronic processes 
initiated by relativistic protons co-accelerated with the electrons 
\citep[for a recent discussion see, e.g.][]{muecke03}. The lack of 
knowledge of the primary jet launching mechanism and the difficulty
in constraining the jet composition from general energetics 
considerations
currently leave both leptonic and hadronic models 
open as viable possibilities. In many cases, both types of models 
can provide acceptable fits to the observed broadband spectral energy
distributions (SEDs) of BL~Lac objects, in particular the TeV blazars
(see, e.g., \cite{mk97,pian98,petry00,kraw02} for leptonic and 
\cite{muecke03} for hadronic models). 

In the framework of leptonic jet models, TeV blazars are successfully
modelled by SSC models in which the high-energy emission is produced
by Compton scattering of electron-synchrotron radiation off the same
ultrarelativistic electrons producing the synchrotron emission
\citep[e.g.][]{mk97,pian98,petry00,kraw02}. Such models have been
successful in modeling not only the SEDs, but also the detailed
spectral variability, including spectral hysteresis at X-ray energies, of 
several TeV blazars \citep[e.g.,][]{krm98,gm98,kataoka00,kusunose00,lk00}.
An inevitable prediction of the SSC model is that any flaring activity
at TeV energies should be accompanied by a quasi-simultaneous flare 
in the synchrotron component. Even if the synchrotron flare does not 
necessarily have to be very pronounced at X-ray energies, since the 
TeV photons might be produced by Compton upscattering of seed photons 
that are observed predominantly in the radio -- optical regime, there
should always be a significant imprint of the TeV flare in the optical
and X-ray light curves.

This prediction is in striking contrast to the recent observation of
\cite{kraw04} of an ``orphan'' TeV flare seen in the Whipple light
curve of the TeV blazar 1ES~1959+650 during a multiwavelength campaign
in the late spring and summer of 2002. The object displayed first a
quasi-simultaneous TeV and X-ray (RXTE) flare, followed by a well
sampled, smooth decline of the X-ray flux over the following 
$\sim 1$~month. However, during this smooth decline, a second 
TeV flare, $\sim 20$~days after the initial one, was observed,
which was only accompanied by very moderate $\lesssim 0.1^{\rm mag}$
flaring activity in the R and V bands. This behavior is clearly
unexpected in a purely leptonic SSC blazar jet model.

In light of their great success to model both the broadband SEDs
and spectral variability of TeV blazars in great detail, leptonic
models might still be a very reasonable starting point for further
investigations of this peculiar flaring behavior of 1ES~1959+650. 
However, even if one assumes that the high-energy emission is 
usually dominated by leptonic processes in blazar jets in general
and in 1ES~1959+650 in particular, one would naturally expect 
that the emitting plasma in blazar jets is not a pure $e^+ e^-$ 
pair plasma, but contains a non-negligible admixture of protons. 
For example, based on X-ray luminosity constraints from observations, 
\cite{sm00} find that even if $e^+ e^-$ pairs outnumber protons 
by a large margin (factor of $\sim 50$), blazar jets might still 
be dynamically dominated by their baryon content. Similar conclusions
have been reached by \cite{kt04}, ruling out a pure electron-proton
plasma in energy equilibrium between electrons and protons or a 
pure electron-positron pair plasma. These conclusions are also
supported by energy requirements in large-scale extragalactic 
X-ray jets observed by {\it Chandra} which seem to remain 
relativistic out to kpc and even Mpc distances from the central 
engine \citep[see, e.g.][]{gc01,sambruna03}).

Detailed simulations of particle acceleration at relativistic
shocks or shear layers show that a wide variety of particle 
spectra may result in such scenarios \citep[e.g.,][]{ob02,so03,ed04},
greatly differing from the standard spectral
index of 2.2 -- 2.3 
previously believed to be a universal
value in relativistic shock 
acceleration \citep[e.g.,][]{gallant99,achterberg01}. Thus, both 
the nature of the matter in blazar jets and the energy spectra of 
ultrarelativistic particles injected into the emission regions in 
blazar jets are difficult to constrain from first principles. 
Consequently, also their kinetic luminosity is hard to constrain. 
However, if Fermi acceleration plays a major role in the energization 
of electrons (pairs) in leptonic jets, then one would naturally expect 
that also protons are accelerated to relativistic energies, though 
conceivably not exceeding the energy threshold to boost the bulk of 
the available soft photons up to the energy of the $\Delta$ resonance at 
1232~MeV in the proton's rest frame to initiate pion production processes. 
While the size-scale constraint would allow the acceleration of protons 
up to Lorentz factors of $\gamma_{\rm p, max} \sim 3 \times 10^8 \, 
B_{-1} \, R_{16}$ (where $B = 0.1 \, B_{-1}$~G is the co-moving 
magnetic field and $R = 10^{16} \, R_{16}$~cm is the size of the 
emitting region), factors related to, e.g., the incomplete development 
of plasma wave turbulences and superluminal magnetic-field configurations 
at oblique shocks \citep{ob02,ed04} may severely limit the maximum energies 
of protons by several orders of magnitude.

It has previously been suggested \citep[e.g.,][]{ad03} that the presence
of external photon fields may substantially lower the effective proton
energy
threshold for $p\gamma$ pion production compared to the
standard hadronic-jet scenario based on synchrotron target photons.
They have also pointed out that the conversion of protons to neutrons
via charged pion production ($p\gamma \to n\pi^+$) may facilitate
the transport of kinetic energy in baryons out to kpc scales. For
$\gamma_{\rm p, max} = 10^4 \, \gamma_4$, photon energies of ${E'}_{\rm ph}
\sim E_{\Delta} / {\gamma'}_{\rm p,max} \sim 30 \, \gamma_4^{-1}$~keV
in the co-moving frame of the emission region would be required in
order to initiate $p\gamma$ processes. Such photon energies are unlikely
to be achieved by intrinsic (electron synchrotron) photons, but they 
may occasionally be provided by external photon sources due to the 
Doppler blue shift into the emission region. For example, quasi-isotropic 
radiation fields from re-processed accretion-disk photons \citep{sbr94,dss97} 
or reflected jet synchrotron emission \citep{gm96,bd98} are good candidates 
for external soft photon sources to occasionally exceed the $p\gamma$ pion 
production threshold for relativistic protons of ${\gamma'}_p \sim 10^3$ 
-- $10^4$. 

This paper presents a discussion of the idea that the ``orphan'' TeV
flare in 1ES~1959+650 resulted from $\pi^0$ decay following $\gamma$p 
pion production on an external photon field dominated by photons from 
the first, simultaneous synchrotron + TeV
flare. 

In \S \ref{model}, the basic model geometry and parameter choices, 
guided by the observations of 1ES~1959+650, are outlined. Analytic 
estimates constraining model parameters, in particular the hadron 
number density and energy content in the jet are presented in 
\S \ref{results}. \S \ref{summary} contains a summary and brief 
discussion. 

\section{\label{model}Model setup and parameter estimates}

\begin{figure}[t]
\includegraphics[height=14cm]{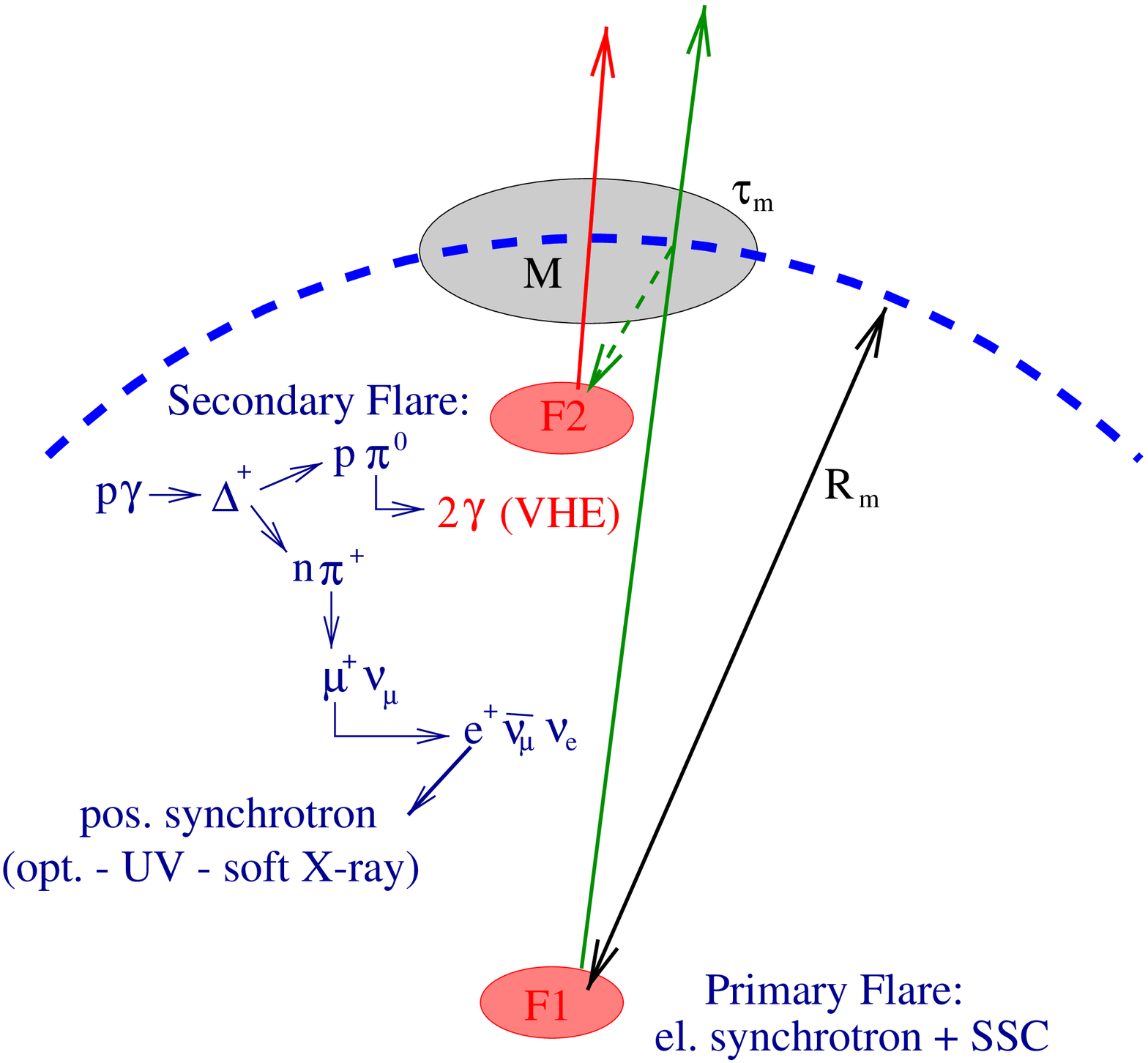}
\caption{Geometry of the model. A primary synchrotron flare is produced
by the emission region (blob) near the center of the system (F1). Synchrotron
emission is reflected at the mirror (M), and re-enters the blob at point
F2, resulting in the secondary, ``orphan'' TeV flare due to $p\gamma$
pion production and subsequent $\pi^0$ decay.}
\label{geometry}
\end{figure}

The basic model geometry is sketched in Fig. \ref{geometry}. A blob filled
with ultrarelativistic electrons and relativistic protons is traveling along
the relativistic jet, defining the positive $z$ axis. Particles are accelerated
very close to the central engine (F1) in an explosive event which is producing 
the initial synchrotron + TeV flare via the leptonic SSC mechanism. Synchrotron
emission from this flare is reflected off a gas cloud (the mirror M) located 
at a distance $R_m$ from the central engine. The cloud has a reprocessing 
optical depth $\tau_m = 10^{-1} \, \tau_{-1}$ and a radius $R_c = 10^{17} \, 
R_{c,17}$~cm, implying an average density of $n_c = 10^6
\, n_6$~cm$^{-3}$
with $n_6 \sim 1.5$.

The characteristic synchrotron photon energy during the primary
flare of 1ES~1959+650 was ${E'}_{\rm sy}
\sim 1 \, \Gamma_1^{-1}
\, E_{\rm sy, 1}$~keV in the co-moving frame, implying a 
characteristic
photon energy of the reflected synchrotron radiation 
of ${E'}_{\rm Rsy}
\sim 100 \, \Gamma_1 \, E_{\rm sy, 1}$~keV, where 
$\Gamma = 10 \, \Gamma_1$ is the bulk
Lorentz factor of the emission 
region, and it is assumed that the
Doppler boosting factor $D \approx 
\Gamma$. Here, the observed peak of the synchrotron spectrum has been 
parametrized as $E_{\rm sy} = 10 \, E_{\rm sy,1}$~keV. Relativistic 
electrons with
${\gamma'}_e \gtrsim 10$ will be very inefficient in 
Compton upscattering this radiation field due to the rapid decline of 
the Klein-Nishina cross section. Here and in the remainder of this paper, 
quantities in the frame of the emission region (``blob'') are denoted 
by primed symbols, while unprimed symbols refer to quantities in the 
stationary system of the AGN. Considering VHE photon production from 
the decay of neutral pions with co-moving Lorentz factors 
${\gamma'}_{\pi^0}$, we may assume that ${\gamma'}_{\pi^0} \approx 
{\gamma'}_{\Delta} \approx {\gamma'}_p$. The observable spectrum of 
$\pi^0$ decay photons will then extend out to $E_{\pi^0 \to 2 \gamma} 
\sim 7 \, \gamma_4 \, \Gamma_1$~TeV.

The observed time delay between the primary synchrotron flare and 
the
secondary flare due to interactions of the blob with the first
reflected
synchrotron flare photons to arrive back at the blob was 
$\Delta t_{\rm obs} = 20 \, \Delta t_{20}$~days, and is related to 
the distance of the reflector by
\begin{equation}
\Delta t_{\rm obs} \approx {R_m \over 2 \, \Gamma^2 c}.
\label{Delta_t}
\end{equation}
Thus, $R_m \approx 3 \Gamma_1^2 \, \Delta t_{20}$~pc. A cloud of 
reflecting
gas with the characteristics specified above, at this 
distance from a central source of the ionizing continuum radiation 
from a central accretion disk with luminosity $L_D = 10^{44} \, 
L_{44}$~ergs~s$^{-1}$ will remain largely  neutral (ionization 
parameter $\xi = L_D / (4 \pi R_m^2 n_c) \sim 8 \times 10^{-2} L_{44} \, 
(R_m / 3 {\rm pc})^{-2} \, n_6^{-1}$). Its optical emission line luminosity
will be limited by $L_{\rm line} < L_D \, (R_c/R_m)^2 = 10^{40} \, 
L_{44} \, R_ {c,17}^2 \, (R_m/3 \, {\rm pc})^{-2}$~ergs~s$^{-1}$,
corresponding to a line flux of $F_{\rm line} < 2 \times 10^{-15} \, 
L_{44} \, R_ {c,17}^2 \, (R_m/3 \, {\rm pc})^{-2}$~ergs~cm$^{-2}$~s$^{-1}$
which is negligible compared to the jet synchrotron continuum, consistent 
with the classification of 1ES~1959+650 as a BL~Lac object. 

\begin{figure}[t]
\includegraphics[height=14cm]{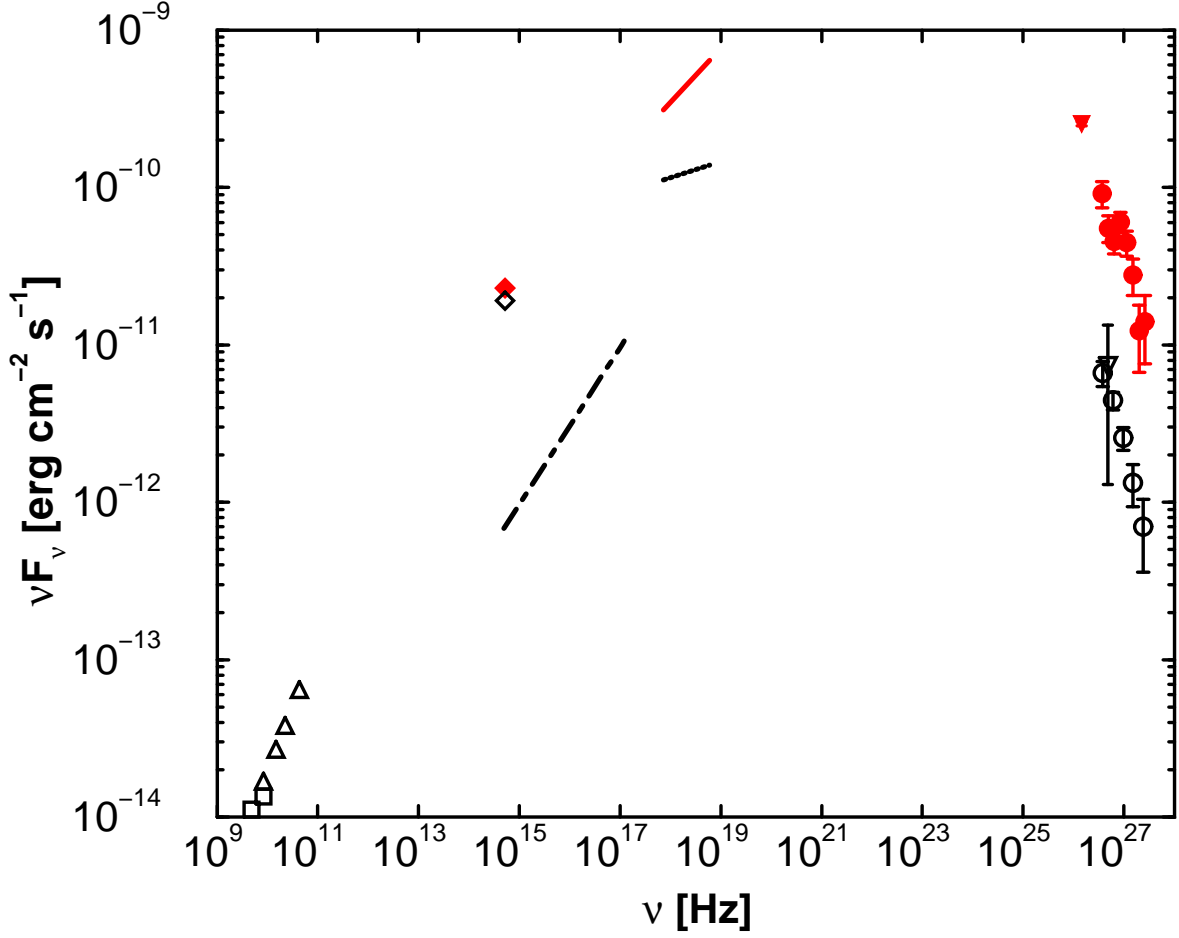}
\caption{Broadband spectral energy distribution of 1ES~1959+650. Filled 
symbols and the solid X-ray spectrum refer to the TeV high state, 
representative of the primary TeV flare; open symbols and the dotted
X-ray spectrum refer to the low TeV state. The dot-dashed power-law
indicates the predicted secondary $e^+$ synchrotron emission following
$\pi^+$ decay. Data from \cite{kraw04}.}
\label{bbspectrum}
\end{figure}

The duration of the flare, $w_{\rm fl}^{\rm obs}$ will then be 
determined by the time it takes for the blob to travel from the 
location $z_0$ of the onset of the secondary flare to the mirror 
at $R_m$:
\begin{equation}
w_{\rm fl}^{\rm obs} = {(R_m - z_0) \, (1 - \beta) \over
\beta \, c} \approx {R_m / 8 \, \Gamma^4 \, c} \approx
1.2 \, \Gamma_1^{-2} \; {\rm hr},
\label{flare_duration}
\end{equation}
where $\beta \, c = \sqrt{1 - 1/\Gamma^2} \, c$ is the speed of 
the
blob which is assumed to remain constant throughout the period 
considered
here. From the observed $\nu F_{\nu}$ fluxes of the 
primary synchrotron
flare and the secondary TeV flare, $\nu F_{\nu} 
({\rm sy}) \sim 5 \times
10^{-10}$~ergs~s$^{-1}$~cm$^{-2}$ and 
$\nu F_{\nu} (600 \, {\rm GeV}) \sim 3 \times 10^{-10}$~ergs~s$^{-1}$~cm$^{-2}$ 
\citep[see Fig. \ref{bbspectrum} and][]{kraw04}, we find the co-moving 
luminosities, ${L'}_{\rm sy} \sim 2.5 \times
10^{41} \, 
\Gamma_1^{-4}$~ergs~s$^{-1}$ and ${L'}_{\rm VHE} \sim 1.5 
\times 10^{41} 
\, \Gamma_1^{-4}$~ergs~s$^{-1}$. Here, an $\Omega_{\Lambda} = 0.7$, 
$\Omega_{\rm m} = 0.3$ cosmology
with $H_0 = 70$~km~s$^{-1}$~Mpc$^{-1}$ 
was used. With these parameters, 1ES~1959+650 with $z = 0.047$ is 
located at a luminosity distance of $d_L = 210$~Mpc. 
If the reflecting cloud is located in 
a direction close
to our line of sight, the energy density of jet 
synchrotron
photons impinging onto the mirror is
\begin{equation}
u_{\rm sy} (R_m) \sim {d_L^2 \over R_m^2 c} \, \nu F_{\nu} ({\rm sy}) 
\sim 1.3 \times 10^5 \, \Gamma_1^{-4} \, \Delta t_{20}^{-2} \; {\rm ergs
\, cm}^{-3}.
\label{u_sy}
\end{equation}
The reflected synchrotron flux will be received by the blob very close
to (and within) the mirror, so that its photon energy density, in the
co-moving frame of the blob, is given by
\begin{equation}
{u'}_{\rm Rsy} \sim {\tau_m \, \Gamma^2 \, u_{\rm sy} (R_m) \over 4 \pi}
\sim 1.0 \times 10^5 \, \Gamma_1^{-2} \, \Delta t_{20}^{-2} \, \tau_{-1}
\; {\rm ergs \, cm}^{-3}.
\label{u_Rsy}
\end{equation}
This reflected synchrotron photon field can now be used to estimate
the energy and density of relativistic protons needed in the jet to
produce the ``orphan'' TeV flare in 1ES~1959+650 via $\gamma$p pion
production and subsequent $\pi^0$ decay, and to estimate the expected
signatures of such a scenario at lower (optical -- X-ray) frequencies.

\section{\label{results}Results}

The co-moving luminosity from $p\gamma \to \Delta \to p+\pi^0 \to 
p +
2 \gamma$ produced by protons of a given energy ${\gamma'}_p$ 
is
given by
\begin{equation}
{L'}_{\rm VHE} \sim {8 \over 3} \, c \, \sigma_{\Delta} \, {u'}_{\rm Rsy}
\gamma'_p {70 \, {\rm MeV} \over {E'}_{\rm Rsy}} \, N_p ({\gamma'}_p).
\label{LRsy}
\end{equation}
where $\sigma_{\Delta} \approx 300 \, \mu$b is the $\Delta$ resonance cross 
section and $N_p ({\gamma'}_p)$ is the number of protons at energy
${\gamma'}_p \approx (300 {\rm MeV})/E'_{\rm Rsy} \approx 3 \times 10^3
\, \Gamma_1^{-1} \, E_{\rm sy,1}^{-1}$. With this, the observable 
$\nu F_{\nu}$ peak flux in the TeV flare can be estimated as
\begin{equation}
\nu F_{\nu} ({\rm VHE}) \sim {{L'}_{\rm VHE} \, \Gamma^4 \over 4 \pi \,
d_L^2} \sim 1.0 \times 10^{-56} \, N_p ({\gamma'}_p) \,
\Delta t_{20}^{-2} 
\, \tau_{-1} \, E_{\rm sy,1}^{-2} \; {\rm ergs \, cm}^{-2}
\, {\rm s}^{-1}.
\label{nFn_VHE}
\end{equation}
Setting this equal to the observed VHE peak flux yields
\begin{equation}
N_p ({\gamma'}_p) \sim 3.0 \times 10^{46} \, \Delta t_{20}^2 \, 
\tau_{-1}^{-1} \, E_{\rm sy,1}^2.
\label{Np_gp}
\end{equation}
The spectrum of non-thermal protons in the blob may be expected
to have a low-energy cut-off at relativistic energies. For example,
if the non-thermal protons are injected into the jet as pick-up
ions from a relativistic shock wave traveling along the jet
\cite[see, e.g.][]{ps00}, this low-energy cutoff is expected
at $\gamma_{\rm p, min} \sim \Gamma = 10 \, \Gamma_1$.
Assuming that the relativistic proton spectrum is a straight 
power-law with index $s$, the estimate (\ref{Np_gp}) corresponds 
to a total relativistic proton number of 
\begin{equation}
N_p \sim
{(3,000)^{1 + s} \cdot 10^{44 - s} \over s - 1} \,
\Gamma_1^{1 - 2 s} \, \Delta t_{20}^2 \, \tau_{-1}^{-1} \,
E_{\rm sy,1}^{2-s}.
\label{Np_total}
\end{equation}
For a typical index $s = 2$, this corresponds to $N_p \sim
2.7 \cdot 10^{52} \, \Gamma_1^{-3} \, \Delta t_{20}^2 \,
\tau_{-1}^{-1}$ and a relativistic proton number density of
\begin{equation}
{n'}_p \sim 6.4 \times 10^3 \, \Gamma_1^{-3} \, \Delta t_{20}^2
\, \tau_{-1}^{-1} \, R_{16}^{-3} \; {\rm cm}^{-3}.
\label{np}
\end{equation}
Note the strong dependence on the bulk Lorentz factor. With 
values of $\Gamma_1 \sim 2$, the required proton density can be 
substantially less than the typical electron densities found 
in spectral
modeling of blazars (${n'}_e \sim 10^3$~cm$^{-3}$), 
which is perfectly consistent with the pair/proton
number density 
ratios inferred by \cite{sm00}. With $s = 2$ and a maximum Lorentz 
factor of relativistic protons of $\gamma_{\rm p, max} \sim 10^4$, 
the total co-moving kinetic energy in
relativistic protons in the 
blob is then
\begin{equation}
{E'}_{b,p} \sim 2.8 \times
10^{51} \, \Gamma_1^{-2} \, 
\Delta t_{20}^2 \, \tau_{-1}^{-1} \; {\rm erg}.
\label{Ep}
\end{equation}
The kinetic luminosity carried by relativistic protons in the 
jet can then be estimated as 
\begin{equation}
L_p^{\rm kin.} \sim
7.3 \times 10^{44} \, R_{16}^{-1} \, 
\Delta t_{20}^2
\, \tau_{-1}^{-1} \, f_{-3} \; {\rm ergs \; s}^{-1}
\label{Lp}
\end{equation}
where $f = 10^{-3} \, f_{-3}$ is a filling factor accounting 
for the likely case that the relativistic proton
plasma is 
concentrated only in individual blobs along the jet rather 
than being continuously distributed throughout the jet.

The radiative output from the $\Delta^+$ decay channel 
$p\gamma \to \Delta^+ \to n\pi^+$, followed by $\pi^+ \to 
\mu^+ + \nu_{\mu}$ and $\mu^+ \to \overline{\nu_{\mu}} + 
e^+ + \nu_e$ will primarily consist of positron synchrotron
radiation. Considering the kinematics of the pion and muon
decay processes, one finds that the positron will carry away
$\sim 1/3$ of the total pion energy. Consequently, we have
$\gamma_{e^+} \sim (1/3) \, (m_{\pi}/m_e) \, \gamma'_p \sim
2.1 \times 10^5 \, \Gamma_1^{-1} \, E_{\rm sy,1}^{-1}$. The
synchrotron emission from the secondary positrons will peak 
at 
\begin{equation}
E_{\rm sy, e^+} \sim 500 \, B_{-1} \, E_{\rm sy,1}^{-2}
\; {\rm eV},
\label{E_psyn}
\end{equation}
i.e. typically in the UV or soft X-ray regime. Note that 
unlike the case of a proton blazar (with higher magnetic 
fields and much higher proton and positron Lorentz factors), 
the charged-pion decay channel will {\it not} initiate an 
electromagnetic cascade. 

An estimate of the expected $\nu F_{\nu}$ flux in the $e^+$ 
synchrotron emission can be found in the following way. First of 
all, considering the co-moving dynamical time scale, $t'_{\rm dyn}
\sim R/c \sim 3.3 \times 10^5 \, R_{16}$~s and the synchrotron
cooling time scale,
\begin{equation}
t'_{e^+ \rm sy} \sim 3.7 \times 10^5 \, \Gamma_1 \,
B_{-1}^{-2} \, E_{\rm sy,1} \; {\rm s},
\label{t_psyn}
\end{equation}
we find that those are comparable, implying that the secondary 
positrons might lose a substantial fraction of their kinetic 
energy to radiation before potentially leaking out of the emission 
region. Second, we realize that the synchrotron cooling time 
scale will set the natural duration of the secondary $e^+$ 
synchrotron flare, which will be (in the observer's frame) 
$w_{e^+ \rm sy}^{\rm obs} \sim 3.7 \times 10^4 \, B_{-1}^{-2} 
\, E_{\rm sy,1}$~s. Consequently, the duration of the $\pi^0$ 
decay VHE $\gamma$-ray flare is a factor of $f_w \equiv 
w_{\pi^0}^{\rm obs} / w_{e^+ \rm sy}^{\rm obs} \sim 0.12 
\, B_{-2}^2 \, E_{\rm sy,1} \, \Gamma_1^{-2}$ shorter than 
the secondary synchrotron flare, so the observed $\nu F_{\nu}$
peak flux of the $e^+$ synchrotron flare should have been a 
factor of $f_w / 3 \sim 0.04$ lower than that of the $\pi^0$ 
decay flare (note that 2/3 of the $\pi^+$ energy will go
into neutrino emission), which yields
\begin{equation} 
\nu F_{\nu}^{e^+ \rm sy} \sim 1.2 \times 10^{-11} \, \Gamma_1^{-2}
\, B_{-2}^2 \, E_{\rm sy,1} \; {\rm ergs \, cm}^{-2} \, {\rm s}^{-1}
\label{nuFnu_psyn}
\end{equation}
if the observed secondary VHE flare was due to $\pi^0$ 
decay photons. The expected $e^+$ synchrotron peak flux
would thus have been only a few \% of the observed RXTE
$\nu F_{\nu}$ flux level during the secondary VHE flare, 
and have peaked at energies well below the RXTE energy 
range, leaving no observable trace in the X-ray light 
curve of the 1ES~1959+650 campaign of 2002. The expected
level and spectral shape of the secondary $e^+$ synchrotron
emission is represented by the dot-dashed line in Fig.
\ref{bbspectrum}.

An estimate of a possible optical flare can be obtained
by realizing that positrons emitting synchrotron radiation
in the optical regime are expected to be slow-cooling and
thus basically reproduce the spectrum of the primary
relativistic protons ($\sim \gamma^{-2}$), resulting in 
a synchrotron spectrum $\nu F_{\nu} \propto \nu^{1/2}$,
which yields an R band flux from the secondary positron
synchrotron emission of
\begin{equation}
\nu F_{\nu}^{e^+ \rm sy} (R) \sim 7.0 \times 10^{-13}
\, \Gamma_1^{-1.5} \, B_{-1}^{1.5} \, E_{\rm sy,1}^2
\; {\rm ergs \, cm}^{-2} \, {\rm s}^{-1}.
\label{nuFnu_R}
\end{equation}
Comparing this to the average R-band flux around the 
time of the secondary (``orphan'') TeV flare, results 
in a predicted optical flare of
\begin{equation}
\Delta m \sim 0.05^{\rm mag},
\label{Delta_m}
\end{equation}
which is perfectly consistent with the observed very
small optical activity of $\Delta m_{\rm obs} \lesssim
0.1^{\rm mag}$ of 1ES~1959+650 at that time \citep{kraw04}.

\section{\label{summary}Summary and Discussion}

In this paper, I have suggested a model to explain the
``orphan'' TeV flare of 1ES~1959+650 in 2002, which followed
a correlated X-ray + TeV flare by about 20~days. In this 
model, the secondary TeV flare resulted from $\pi^0$ decay
following p$\gamma$ pion production by relativistic protons
($\gamma_p \sim 10^3$ -- $10^4$) on the primary synchrotron
flare photons, reflected off a mirror cloud at a distance
of a few pc from the central engine. Using the observational
data from the 1ES~1959+650 observations in 2002, I have 
estimated the required parameters pertaining to the relativistic
proton population in the jet in order to produce the secondary
TeV flare with this mechanism. The main results of this investigation 
are:

\begin{itemize}

\item The required model setup is consistent with the BL~Lac
classification of 1ES~1959+650.

\item The required density of relativistic protons in the jet
is very well consistent with earlier findings that blazar jets
might be dynamically dominated by the kinetic energy of relativistic
protons, even if they are by far outnumbered by electron/positron
pairs, which may dominate the radiative output of 1ES~1959+650
most of the time.

\item The secondary $e^+$ synchrotron emission resulting from
$\pi^+$ decay in this scenario is too weak and peaks at too low
energies to leave an observable imprint in the RXTE light curve
at the time of the secondary TeV flare, consistent with its 
non-detection (and, thus, with the appearance of the TeV flare
as an ``orphan'' flare).

\item The optical flare produced by secondary $e^+$ synchrotron
emission is expected to produce only a very mild bump of
$\Delta m \sim 0.05^{\rm mag}$ in the R and V bands, which
is perfectly consistent with the very moderate activity of
the source during the secondary TeV flare.

\end{itemize}

A detailed investigation of the spectral and light curve
features resulting in this scenario is currently underway
and will be published in a forthcoming paper (Postnikov \&
B\"ottcher 2004, in preparation). This will also include
the characteristics of the expected neutrino emission
resulting from $\pi^+$ decay.

Another signature of relativistic protons in the framework
of the model suggested here might arise from photo-pair
production, $p\gamma \to p e^+ e^-$. The threshold proton
energy for this process, in our parametrization is $\gamma_{\rm
thr, pair} \sim 5 \, \Gamma^{-1} \, E_{\rm sy, 1}^{-1}$. The
bulk of pairs injected into the emission region from this process
would thus have only mildly relativistic energies and would not
leave significant non-thermal radiation signatures. However, it
has been demonstrated by \cite{km99} \cite[see also][for the
application of this process to gamma-ray bursts]{kgm02,kgm04} 
that the photo-pair production process can exceed a critical 
threshold beyond which a pair avalanche on synchrotron radiation
of secondary pairs develops. The threshold proton energy to
initiate such an avalanche has been evaluated by \cite{km99}
to be $\gamma_{\rm p, crit} \sim 10^4 \, B_{\rm G}^{-1/3}
\, \Gamma_1^{-2/3}$. Thus, if the emitting volume contains 
protons with energies $\gamma_p \gg 10^4$, this supercritical
pair avalanche can lead to a strong synchrotron signal, extending
far into the X-ray regime, which would naturally be expected to
produce a corresponding SSC signature at $\gamma$-ray energies. 
Because of the strong synchrotron component of this scenario, 
these signatures are easily distinguishable. It is very well
conceivable that the ``supercritical pile'' scenario \citep{km99},
indicative of protons with Lorentz factors of $\gamma_p \gg 10^4$,
is responsible for simultaneous X-ray + TeV $\gamma$-ray flares,
while the pion production scenario discussed here, indicative of
protons with Lorentz factors of $10^3 \lesssim \gamma_p \lesssim
10^4$ produces orphan TeV flares. Protons with yet lower Lorentz
factors may ultimately be probed by the radiation signatures of
mildly relativistic or even thermal pairs injected through the
$p\gamma$ pair production process of protons near threshold.

\acknowledgments
The author wishes to thank H. Krawczynski for providing the 
broadband spectral data of 1ES~1959+650, and the anonymous
referee for pointing out the potential importance of the
$p\gamma$ process in the framework of this model. This work 
was partially supported by NASA through INTEGRAL GO grant 
no. NAG~5-13684 and XMM-Newton GO grant no. NNG~04GF70G.

\newpage

\newpage

\end{document}